\documentclass[conference]{IEEEtran}
\usepackage{amsmath,amsopn,amssymb,bm}
\usepackage{amsfonts}
\usepackage{amsthm} 
\usepackage{graphicx,xspace,color}
\usepackage{epsfig}
\usepackage{longtable,multirow}
\usepackage{array}
\usepackage{stfloats}
\usepackage{cuted}
\usepackage{cite}
\usepackage[nolist]{acronym}
\usepackage{soul}
\usepackage{algorithm}
\usepackage[noend]{algpseudocode}
\usepackage{algcompatible}
\usepackage{enumerate}
\usepackage{lettrine}
\usepackage{mathtools}
\usepackage{comment}
\usepackage{subcaption}
\usepackage{romannum}

\usepackage[font=small]{caption}
\usepackage{booktabs}

\graphicspath{ {./figures/}}

\newtheorem{remark}{Remark}

\begin{document}

\begin{acronym}
	\acro{SCD}{sinusoidal current distribution}
	\acro{EE}{energy efficiency}
	\acro{MIMO}{Multiple-input multiple-output}
	\acro{SNR}{signal-to-noise ratio}
	\acro{NPA}{non-uniform planar array}
	\acro{ULA}{uniform linear array}
	\acro{MRT}{maximum ratio transmission}
	\acro{Rx}{receiver}
	\acro{IEs}{integral equations}
	\acro{MoM}{method of moments}
\end{acronym}

\title{Superdirective Arrays with Finite-Length Dipoles: Modeling and New Perspectives}

\author{\IEEEauthorblockN{Konstantinos Dovelos, Stylianos D. Assimonis, Hien Quoc Ngo, and Michail Matthaiou\\}
	\IEEEauthorblockA{Centre for Wireless Innovation, Queen's University Belfast, Belfast, U.K.\\}
	Email: \{k.dovelos, s.assimonis, hien.ngo, m.matthaiou\}@qub.ac.uk
}

\maketitle

\begin{abstract}
Dense arrays can facilitate the integration of multiple antennas into finite volumes. In addition to the compact size, sub-wavelength spacing enables superdirectivity for endfire operation, a phenomenon that has been mainly studied for isotropic and infinitesimal radiators. In this work, we focus on linear dipoles of arbitrary yet finite length. Specifically, we first introduce an array model that accounts for the \ac{SCD} on very thin dipoles. Based on the \ac{SCD}, the loss resistance of each dipole antenna is precisely determined. Capitalizing on the derived model, we next investigate the maximum achievable rate under a fixed power constraint. The optimal design entails conjugate power matching along with maximizing the array gain. Our theoretical analysis is corroborated by the method of moments under the thin-wire approximation, as well as by full-wave simulations. Numerical results showcase that a super-gain is attainable with high radiation efficiency when the dipole antennas are not too short and thin.

\end{abstract}

\section{Introduction}
\ac{MIMO} systems have shaped modern wireless communications thanks to their unique capabilities, ranging from spatial multiplexing to sharp beamforming~\cite{prospects_mimo}. However, deploying a massive antenna array entails several challenges, such as high power consumption and size. To this end, compact arrays with sub-wavelength spacing emerge as a promising solution for beyond massive \ac{MIMO} communication~\cite{beyond_mimo}. In addition to the small footprint of dense arrays, extremely large power gains can be attained by exploiting the mutual coupling of closely spaced antennas, a concept known as \textit{superdirectivity}. Specifically, Uzkov~\cite{uzkov} theoretically showed that for a \ac{ULA} with $N$ isotropic elements and a vanishingly small interelement spacing, the maximum array directivity approaches~$N^2$. This astonishing theoretical result has ignited a great research interest in the fundamental limits of phased arrays since then. 

On the negative side, it is known that superdirectivity requires high antenna currents, which can undermine its implementation in practice~\cite{theory_linear_arrays}. This problem is exacerbated when employing a large number of antenna elements. A stream of prominent papers (e.g.,~\cite{superdirectivity_mimo_loss,superdirectivity_mimo,marzetta_1,marzetta_2,marzetta_3,he_sg_arrays,circuit_th_commun}, and references therein) investigated the performance of dense antenna arrays, yet considering rather simplistic antenna models. In particular, they assumed either isotropic radiators or Hertzian dipoles. However, the latter have infinitely large input reactance; thus, impedance matching is impossible as highlighted also in~\cite{marzetta_3}. Moreover, electrically small antennas suffer from poor radiation efficiency in general. From the related literature, we distinguish~\cite{nf_mimo} which studied near-field \ac{MIMO} communication with half-wavelength dipoles. Yet, existing works on superdirectivity overlook the physical dimensions of the array elements, which can have great impact on the radiation efficiency of the system. In this paper, we aim to fill this gap in the literature and shed light on the fundamentals of superdirectivity with linear dipoles. The contributions of the paper are summarized as follows:
\\[-0.47cm]
\begin{itemize}
\item We provide an electromagnetic model for arrays of dipoles with arbitrary length. To facilitate analysis, a \acf{SCD}~\cite{balanis_book} is assumed on each dipole. Leveraging the \ac{SCD}, the loss resistance of each dipole antenna is analytically determined. Note that ohmic losses play a key role in the performance of superdirectivity~\cite[Ch. 6]{balanis_book}, and hence their proper modeling is of the utmost importance. 
\item Building upon the derived array model, we study the achievable rate under a fixed power constraint. In particular, the optimal design entails single-port power matching based on the notion of active impedance, which eliminates reflection losses; thus, it guarantees maximal power transfer between the voltage sources and the antenna elements in the presence of mutual coupling. Furthermore, beamforming is performed by maximizing the array gain. In this way, a super-gain is attained whilst increasing the energy efficiency of the system.
\item Since the \ac{SCD} assumption is accurate for infinitely thin wires, we validate our theoretical findings by the \ac{MoM} and full-wave simulations with 4NEC2~\cite{mom_book}. For the \ac{MoM}, a comprehensive framework relying on the antenna currents obtained by Hall\'{e}n's \ac{IEs} is presented. It is then shown that the \ac{SCD}-based model produces accurate results for coupled dipoles of finite radius. Consequently, the proposed model can be used to theoretically study superdirectivity without resorting to cumbersome full-wave simulations.
\item Our analysis reveals the interplay between dipoles' dimensions and superdirectivity. Particularly, it is demonstrated that increasing the dipoles' length to specific values yields higher array gain with smaller antenna currents than short antennas. This novel observation can facilitate the efficient implementation of superdirective arrays for beyond 5G applications, ranging from wireless power transfer to nonterrestrial communications.
\end{itemize}
\textit{Notation}: $\mathbf{a}$ is a vector; $\mathbf{A}$ is a matrix; $[\mathbf{A}]_{i,j}$ is the $(i,j)$th entry of $\mathbf{A}$; $(\cdot)^T$, $(\cdot)^*$, and $(\cdot)^H$ denote the transpose, conjugate, and conjugate transpose, respectively; $\|\mathbf{a}\|$ is the $l_2$-norm of $\mathbf{a}$; $\mathbf{a}\cdot \mathbf{b}$ is the inner product between $\mathbf{a}$ and $\mathbf{b}$; $\mathbf{I}_N$ is the $N\times N$ identity matrix; and $\text{Re}\{\cdot\}$ is the real part of a complex variable.

\section{Model of Dipole Array}\label{sec:system_model}
In this section, we propose an array model for lossy antennas based on electromagnetic theory. Consider an array of $N$ linear dipoles, each having length~$\ell$ and radius~$\rho$. All dipoles are parallel to the $z$-axis and are center-fed by voltage sources which induce antenna currents. We next assume that the current distribution on each dipole $n$ has approximately the form~\cite[Ch. 4]{balanis_book}
\begin{equation}\label{eq:sin_current_approximation}
I_n(z')  \approx I_n(0)\frac{\sin\left(\frac{k\ell}{2}-k|z'|\right)}{\sin\left(\frac{k\ell}{2}\right)}, \quad |z'| \leq \ell/2,
\end{equation}
where $I_n(0)\in\mathbb{C}$ is the input current, $k = 2\pi/\lambda$ is the wavenumber, and $\lambda$ is the carrier wavelength.

\subsection{Radiated Power}
Let $(r\cos\phi\sin\theta,r\sin\phi\sin\theta,r\cos\theta)$ be the \ac{Rx} location, where $r$, $\phi\in[0,2\pi]$, and $\theta\in [0,\pi]$ are the radial distance, azimuth angle, and polar angle, respectively. The \ac{Rx} is in the far-field zone of the antenna array. The magnitude of the electric field at the \ac{Rx} is then specified as~\cite[Ch. 4]{balanis_book}
\begin{align}\label{eq:e_field}
E_\theta = \frac{ jZ_0 e^{-jkr}}{2\pi r}\frac{\cos\left( \frac{k\ell}{2}\cos\theta\right)\!-\cos\left( \frac{k\ell}{2}\right)}{\sin\left(\frac{k\ell}{2}\right)\sin\theta} \sum_{n=0}^{N-1}\! e^{jk \hat{\mathbf{r}}\cdot \mathbf{r}_n} I_n(0),
\end{align} 
where $Z_0$ denotes the characteristic impedance of free-space, $\hat{\mathbf{r}} = (\cos\phi\sin\theta,\sin\phi\sin\theta,\cos\theta)^T$ is the unit radial vector along the \ac{Rx} direction, and $\mathbf{r}_n\in\mathbb{R}^{3\times 1}$ is the position vector of the $n$th antenna. The radiation intensity [W/sr] is written in vector form as
\begin{equation}\label{radiation_intensity}
U \triangleq \frac{|E_\theta|^2}{2Z_0} r^2= \frac{Z_0}{8\pi^2}F^2(\theta)\left|\mathbf{a}^H(\theta,\phi)\mathbf{i}\right|^2,
\end{equation}
where $F(\theta) = [\cos(k \ell/2\cos\theta)-\cos(k\ell/2)]/[\sin(k\ell/2)\sin\theta]$ corresponds to the field pattern of an individual dipole, $\mathbf{i} = [I_0(0),\dots, I_{N-1}(0)]^T\in \mathbb{C}^{N\times 1}$ is the vector of input currents, and $\mathbf{a}(\theta,\phi) = [e^{-jk \hat{\mathbf{r}}\cdot \mathbf{r}_0},\dots,e^{-jk \hat{\mathbf{r}}\cdot \mathbf{r}_{N-1}}]^T\in\mathbb{C}^{N\times 1}$ is the array response vector. Using~\eqref{radiation_intensity}, the power radiated by the antenna array is 
\begin{align}\label{radiated_power_em}
P_{\text{rad}} &= \int_0^{\pi}\!  \int_0^{2\pi} U \sin\theta \text{d}\theta \text{d}\phi  = \frac{1}{2}\mathbf{i}^H\mathbf{Z}_{\text{real}}\mathbf{i},
\end{align}
where $\mathbf{Z}_{\text{real}}\in\mathbb{R}^{N\times N}$ is the real-valued matrix with entries
\begin{equation}\label{eq:z_real_elements}
[\mathbf{Z}_{\text{real}}]_{n,m} = \frac{Z_0}{4\pi^2}\int_0^{\pi}\!  \int_0^{2\pi}\! \! \! \! e^{-jk \hat{\mathbf{r}}\cdot (\mathbf{r}_n-\mathbf{r}_m)}F^2(\theta)\sin\theta \text{d}\theta \text{d}\phi. \\[0.1cm]
\end{equation}

\begin{remark}
Expression~\eqref{eq:e_field} relies on the pattern multiplication principle, whereby the electric field is the product of the array factor $\mathbf{a}^H\mathbf{i}$ and the field pattern $F(\theta)$ of an isolated dipole radiating in free-space. This implies that the current distribution on each dipole is not affected by the presence of other antennas, and hence can be considered as sinusoidal. The accuracy of this postulate is further examined in Section~\ref{sec:model_valication}.
\end{remark}

\subsection{Input Power and Array Gain} Realistic antennas exhibit a loss resistance which leads to heat dissipation. Because of the \textit{skin effect} of conductive wires carrying an alternating current, the loss resistance per unit length is given by~\cite[Eq. (2-90b)]{balanis_book}
\begin{equation}\label{eq:loss_resistance_pul}
\bar{R}_{\text{loss}} = \frac{1}{2\rho}\sqrt{\frac{f\mu}{\pi\sigma}},
\end{equation}
where $f$ is the carrier frequency, $\mu$ is the permeability of free-space, and $\sigma$ is the conductivity of the wire material. Under the \ac{SCD} in~\eqref{eq:sin_current_approximation}, the loss resistance relative to the input current $I_n(0)$ is given by
\begin{align}\label{eq:loss_resistance}
R_{\text{loss}} &= \bar{R}_{\text{loss}}
\int_{-\ell/2}^{
	\ell/2} \left|\frac{I_n(z')}{I_n(0)}\right|^2 \text{d}z'=  \frac{k  \ell - \sin(k\ell)}{4 k \rho\sin^2\left(\frac{k\ell}{2}\right)}\sqrt{\frac{f\mu}{\pi\sigma}}, 
\end{align}
which yields the overall power loss 
\begin{equation}
P_{\text{loss}} = \frac{1}{2}\sum_{n=0}^{N-1}R_{\text{loss}}|I_n(0)|^2 = \frac{1}{2}R_{\text{loss}}\|\mathbf{i}\|^2.
\end{equation}
As a result, the input power at the antenna ports is 
\begin{align}
P_{\text{in}} = P_{\text{loss}} + P_{\text{rad}} &= \frac{1}{2}\mathbf{i}^H(R_{\text{loss}}\mathbf{I}_N + \mathbf{Z}_{\text{real}})\mathbf{i} \nonumber \\
& = \frac{1}{2}\mathbf{i}^H\text{Re}\{\mathbf{Z}_{\text{in}}\}\mathbf{i},
\end{align}
where $\mathbf{Z}_{\text{in}}\triangleq R_{\text{loss}}\mathbf{I}_N + \mathbf{Z}$ is the input impedance matrix of the array; $\mathbf{Z}\in\mathbb{C}^{N\times N}$, with $\text{Re}\{\mathbf{Z}\} = \mathbf{Z}_{\text{real}}$, is the input impedance matrix for lossless antennas. Finally, the array gain is defined~as
\begin{equation}\label{eq:gain_sin}
G(\theta,\phi)   \triangleq \frac{4\pi U }{P_{\text{in}}} =   \frac{Z_0F^2(\theta)}{\pi}\frac{|\mathbf{a}^H(\theta,\phi)\mathbf{i}|^2}{\mathbf{i}^H\text{Re}\{\mathbf{Z}_{\text{in}}\}\mathbf{i}},
\end{equation}
and the power at the \ac{Rx} is determined as
\begin{equation}
P_r = P_{\text{in}}\left(\frac{\lambda }{4\pi r}\right)^2 G(\theta,\phi) ,
\end{equation}
where an isotropic receiving antenna has been assumed. 

\subsection{Total Power and Matching Efficiency }
In a practical scenario, the dipoles are driven by voltage sources. To this end, we consider that a voltage source is connected to each antenna port $n$ through the impedance $Z_{M,n}$ used for single-port power matching.\footnote{Multi-port matching requires inter-connections across all antenna ports. Thus, it can become very complicated in massive antenna arrays~\cite{matching_strategies,Nossek_1}.} The total power consumption of the array is now determined as
\begin{align}
P_{\text{total}} &=  \frac{1}{2}\mathbf{i}^H\text{Re}\{\mathbf{Z}_M\}\mathbf{i} + P_{\text{in}}\nonumber \\
&= \frac{1}{2} \mathbf{i}^H(\text{Re}\{\mathbf{Z}_M\} + \text{Re}\{\mathbf{Z}_{\text{in}}\})\mathbf{i},
\end{align}
where $\mathbf{Z}_M = \text{diag}(Z_{M,0},\dots,Z_{M,N-1})$. For a given $P_{\text{total}}$, the received power is finally recast as
\begin{equation}
P_r = \eta P_{\text{total}}\left(\frac{\lambda }{4\pi r}\right)^2 G(\theta,\phi) ,
\end{equation}
where $\eta \triangleq P_{\text{in}}/P_{\text{total}}\in [0,1/2]$ is the matching efficiency accounting for potential reflection losses due to impedance mismatch. With perfect matching, $\eta = 1/2$, which implies that half of the total power is delivered to the antenna array~\cite{pozar_book}.
\begin{figure*}[t]
	\centering
	\begin{subfigure}{\textwidth}
		\centering
		\includegraphics[width=0.95\linewidth]{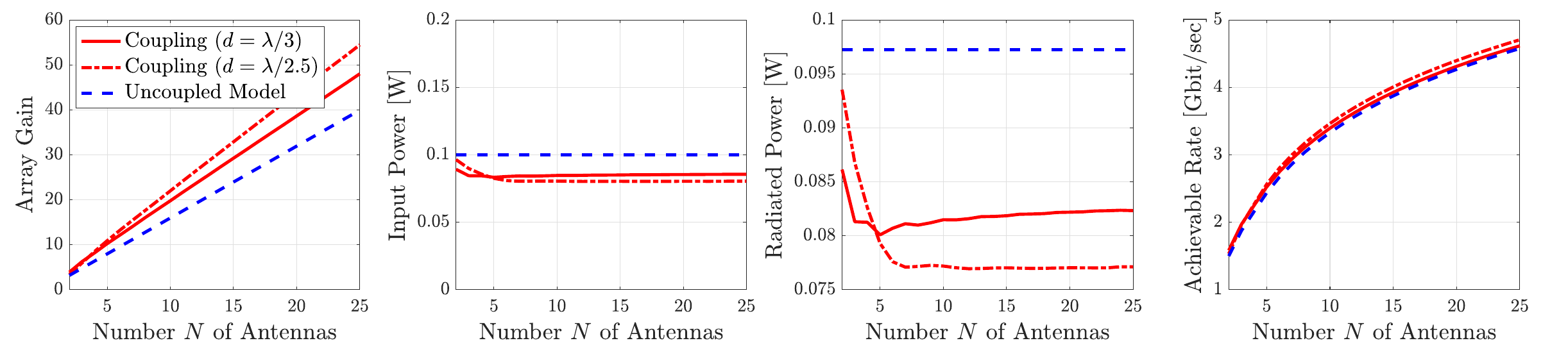}
		\caption{\footnotesize Conjugate matching with input impedance: $Z_{M,n} = [\mathbf{Z}_{\text{in}}]^*_{n,n}$.}
		\label{fig:Fig1a}
	\end{subfigure}
	\begin{subfigure}{\textwidth}
		\centering
		\includegraphics[width=0.95\linewidth]{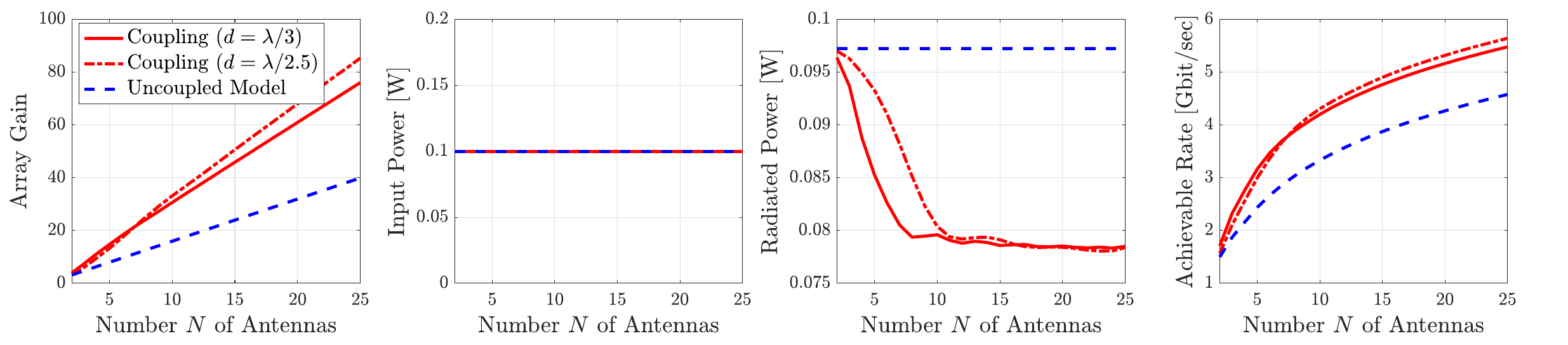}
		\caption{\footnotesize Conjugate matching with active impedance: $Z_{M,n} = [\mathbf{Z}_a]^*_{n,n}$.}
		\label{fig:Fig1b}
	\end{subfigure}%
	\caption{Results versus number of antennas for endfire \ac{ULA} with interelement spacing $d$. The elements have $\ell = \lambda/2$ and $\rho = \lambda/2000$, are made of copper with conductivity $\sigma = 5.7 \times 10^7$ S/m, and are placed along the $x$-axis, i.e., $\mathbf{r}_n = (nd,0,0)$. The \ac{Rx} is at $r= 500$~m and $(\theta,\phi) = (\pi/2,0)$. The other parameters are: $f=10$ GHz, $W= 1$ GHz, $P_t = 200$ mW, and $\sigma_n^2= -174$ dBm/Hz.}
	\label{fig:Fig1}
\end{figure*}

\section{Optimal Design under Mutual Coupling}
\subsection{Beamforming}
The achievable rate [\text{bit/sec}] is specified as
\begin{align}
R &=  W\log_2\left(1 + \frac{P_r}{W\sigma_n^2}\right) \nonumber \\
& =  W\log_2\left(1 + \frac{P_{\text{total}}}{W\sigma_n^2}\frac{\lambda^2}{(4\pi r)^2} \eta G(\theta,\phi) \right),
\end{align}
where $W$ is the signal bandwidth, and $\sigma^2_n$ is the noise power density at the \ac{Rx}. We next seek to find $\mathbf{i}$ that maximizes~$R$ under the constraint $P_{\text{total}}\leq P_t$, where $P_t$ denotes the maximum power budget of the system. By properly scaling the vector $\mathbf{i}$ of currents, $P_{\text{total}} = P_t$ and $\eta G(\theta,\phi) $ remains unchanged. Then, the initial problem of maximizing $R$ becomes equivalent to the unconstrained problem
\begin{align}\label{eq:opt_problem_gain}
\max_{\mathbf{i}} \ \eta G(\theta,\phi)  =  \frac{\mathbf{i}^H\mathbf{a}(\theta,\phi)\mathbf{a}^H(\theta,\phi)\mathbf{i}}{\mathbf{i}^H(\text{Re}\{\mathbf{Z}_M\} + \text{Re}\{\mathbf{Z}_{\text{in}}\})\mathbf{i}}.  
\end{align}
The objective in~\eqref{eq:opt_problem_gain} is a generalized Rayleigh quotient, and hence it admits the solution $\mathbf{i} = \mathbf{C}^{-1}\mathbf{a}(\theta,\phi)$, where $\mathbf{C} \triangleq \text{Re}\{\mathbf{Z}_M\} + \text{Re}\{\mathbf{Z}_{\text{in}}\}$ for notational convenience. Given the above, the optimal current vector is 
\begin{equation}\label{eq:optimal_bf}
\mathbf{i} =\sqrt{\frac{ 2P_t}{\mathbf{a}^H(\theta,\phi)\mathbf{C}^{-1}\mathbf{a}(\theta,\phi)}}\mathbf{C}^{-1}\mathbf{a}(\theta,\phi).
\end{equation}

\subsection{Single-Port Matching}
Mutual coupling alters the input impedance of each dipole. Thus, typical conjugate matching, i.e., $Z_{M,n} = [\mathbf{Z}_{\text{in}}]^*_{n,n}$, will result in significant reflection losses. To avoid this, we leverage the notion of \textit{active impedance}, which follows from the relationship $\mathbf{v} = \mathbf{Z}_{\text{in}}\mathbf{i} = \mathbf{Z}_a\mathbf{i}$, where $\mathbf{v}\in\mathbb{C}^{N\times 1}$ is the vector of voltages at the antenna ports. The active impedance matrix $\mathbf{Z}_a\in\mathbb{C}^{N\times N}$ is diagonal with entries
\begin{equation}\label{eq:active_impedance}
[\mathbf{Z}_a]_{n,n}= \left(R_{\text{loss}} + [\mathbf{Z}]_{n,n} + \sum_{m=0, m \neq n}^{N-1}[\mathbf{Z}]_{n,m}\frac{i_m}{i_n}\right).
\end{equation}
The reflection coefficient for the $n$th port is defined as~\cite{elect_small_antenna}
\begin{equation}\label{eq:reflect_coeff}
\Gamma_n \triangleq \frac{[\mathbf{Z}_a]_{n,n} - Z^*_{M,n}}{[\mathbf{Z}_a]_{n,n} + Z_{M,n}}.
\end{equation}
From~\eqref{eq:reflect_coeff}, it is evident that optimal matching is accomplished for $Z_{M,n} = [\mathbf{Z}_a]^*_{n,n}$. Note that the active impedance matrix hinges on the vector $\mathbf{i}$ of currents, and hence it changes with $(\theta,\phi)$. Consequently, reflectionless operation is possible only for a specific scanning direction $(\theta,\phi)$. The entries of the impedance matrix $\mathbf{Z}$ used in~\eqref{eq:active_impedance} are calculated by the induced EMF method~\cite[Ch. 25]{orfanidis_book}. Under the optimal matching strategy, the total power becomes
\begin{align}\label{eq:total_power_pm}
P_{\text{total}} &= \frac{1}{2}\text{Re}\left\{\mathbf{i}^H(\mathbf{Z}^*_a + \mathbf{Z}_{\text{in}})\mathbf{i}\right\} \nonumber \\
& = \mathbf{i}^H\text{Re}\{\mathbf{Z}_{\text{in}}\}\mathbf{i},
\end{align}
which is exactly twice the input power. For this reason, the beamforming problem~$\eqref{eq:opt_problem_gain}$ reduces to maximizing the array gain, i.e., $\max_{\mathbf{i}} \ G(\theta,\phi) /2 = \max_{\mathbf{i}} \  G(\theta,\phi) $. Given that, the maximum array gain at the \ac{Rx} direction $(\theta,\phi)$ is
\begin{equation}
G_{\max}(\theta,\phi) = 	\frac{Z_0F^2(\theta)}{\pi}\mathbf{a}^H(\theta,\phi)\text{Re}\{\mathbf{Z}_{\text{in}}\}^{-1}\mathbf{a}(\theta,\phi). 
\end{equation}
\begin{figure*}[t]
	\centering
	\begin{subfigure}{\textwidth}
		\centering
		\includegraphics[width=0.95\linewidth]{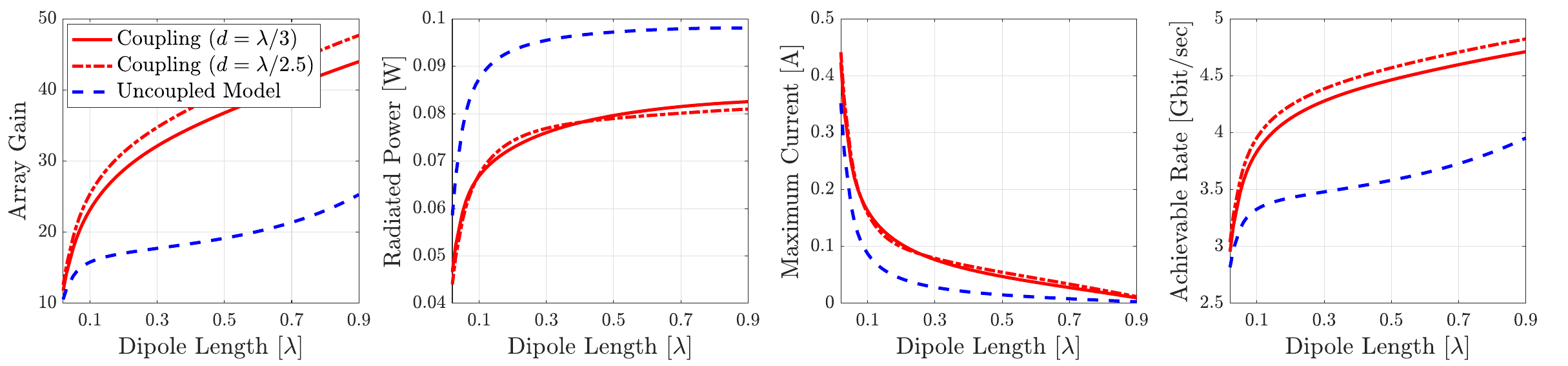}
		\caption{\footnotesize $\rho = \lambda/2000$}
		\label{fig:Fig2a}
	\end{subfigure}
	\begin{subfigure}{\textwidth}
		\centering
		\includegraphics[width=0.95\linewidth]{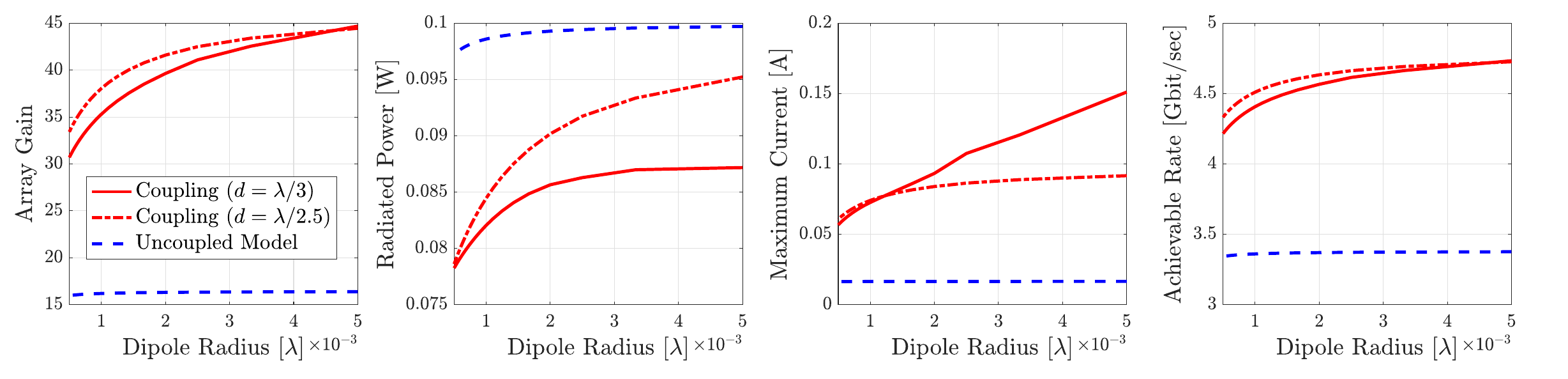}
		\caption{\footnotesize $\ell = \lambda/2$}
		\label{fig:Fig2b}
	\end{subfigure}%
	\caption{Results for endfire \ac{ULA} with $N=10$ elements and interlement spacing $d$. The elements are made of copper with conductivity $\sigma = 5.7 \times 10^7$ S/m, and are placed along the $x$-axis, i.e., $\mathbf{r}_n = (nd,0,0)$. The \ac{Rx} is at $r= 500$~m and $(\theta,\phi) = (\pi/2,0)$. The other parameters are: $f=10$ GHz, $W= 1$ GHz, $P_t = 200$ mW, and $\sigma_n^2= -174$ dBm/Hz.}
	\label{fig:Fig2}
\end{figure*}
\\[-1cm]
\subsection{Uncoupled Model}
In the absence of mutual coupling, $\mathbf{Z}_{\text{real}} = R_i\mathbf{I}_N$, where $R_i \triangleq[\mathbf{Z}_{\text{real}}]_{n,n}$ defines the input resistance of a lossless dipole, i.e., radiation resistance divided by $\sin^2(k\ell/2)$~\cite[Ch.~8]{balanis_book}. Then, $P_{\text{rad}} =\frac{1}{2}R_i\|\mathbf{i}\|^2$, which is exactly the power emitted by $N$ uncoupled antennas. Moreover,~\eqref{eq:optimal_bf} reduces to
\begin{equation}\label{eq:current_unc}
\mathbf{i} = \sqrt{\frac{P_t}{N(R_\text{loss}+ R_i)}}\mathbf{a}(\theta,\phi).
\end{equation}
Under~\eqref{eq:current_unc}, $G_{\max}(\theta) = \frac{Z_0}{\pi (R_{\text{loss}}+R_i)} F^2(\theta) N = O(N)$, which is the conventional power gain that increases linearly with the number $N$ of antennas.
\section{Numerical Results and Discussion}
\subsection{Performance versus Number of Antennas}
In this numerical experiment, we consider half-wavelength dipoles. From Fig.~\ref{fig:Fig1}, we first observe the superdirectivity effect thanks to strong mutual coupling. The importance of suitable impedance matching is also showcased in Fig.~\ref{fig:Fig1}(\subref{fig:Fig1a}), where reflection losses cancel out the benefit of superdirectivity. It is worth stressing that the array gain is reduced when $Z_{M,n} = [\mathbf{Z}_{\text{in}}]_{n,n}^*$ because the optimal excitation \eqref{eq:optimal_bf} maximizes the product $\eta G(\theta,\phi)$. 

Under perfect matching, the reduction in the radiated power due to heat dissipation is compensated by the large increase in the directivity. As a result, the achievable rate is significantly enhanced by employing a sub-wavelength spacing, whilst meeting the power constraint $P_{\text{total}} \leq 200$ mW. In short, superdirectivity does not necessarily compromise the energy efficiency of the system. Regarding the uncoupled case, the radiated power and ohmic losses remain constant versus $N$ as
\begin{align}\label{eq:p_loss_unc}
P_{\text{rad}} &= \frac{1}{2}\frac{R_i}{R_{\text{loss}} + R_i}P_t,  \\
P_{\text{loss}}  &= \frac{1}{2}\frac{R_{\text{loss}}}{R_{\text{loss}} + R_i}P_t.
\end{align}
This comes in sharp contrast to the superdirective case, where ohmic losses become dominant for a large number of antennas. To mitigate this problem, one can adopt longer dipoles to boost the transmission efficiency of each array element.

\subsection{Effect of Dipole Dimensions}
From antenna theory, we know that longer dipoles have higher element directivity and radiation resistance~\cite{antenna_book}. Hence, they can be beneficial in terms of transmission characteristics. Furthermore, the elements of $\mathbf{Z}_{\text{in}}$ become larger for $\ell > \lambda/2$, which results in smaller antenna current values as $\mathbf{i}\propto \mathbf{Z}^{-1}_{\text{in}}\mathbf{a}$. This behavior becomes more evident in the uncoupled case, where $\mathbf{i}\propto (R_\text{loss}+ R_i)^{-1/2}$. This finding is validated in Fig.~\ref{fig:Fig2}(\subref{fig:Fig2a}) for $0.02\lambda \leq \ell \leq 0.9\lambda$. In short, increasing the dipole length up to $0.9\lambda$ improves the radiation efficiency of superdirectivity. Regarding dipoles' radius,~\eqref{eq:loss_resistance} shows that the loss resistance is inversely proportional to $\rho$. Consequently, increasing the dipoles' radius will decrease the ohmic losses of the array. Figure~\ref{fig:Fig2}(\subref{fig:Fig2b}) demonstrates the benefit of employing thicker dipoles for $\lambda/2000 \leq \rho \leq \lambda/200$.  
\begin{figure*}[t]
	\centering
	\begin{subfigure}{.3\textwidth}
		\centering
		\includegraphics[width=0.78\linewidth]{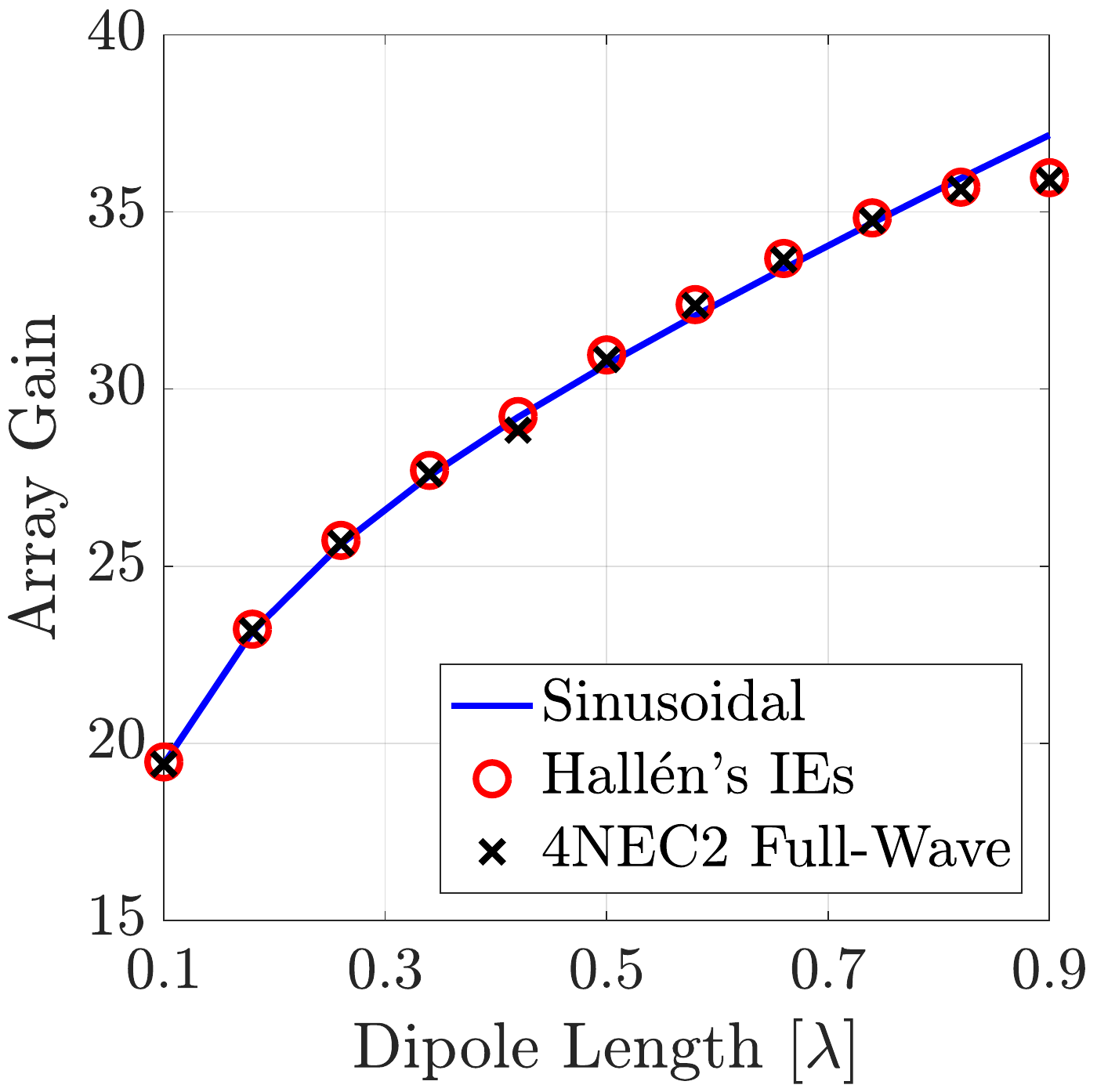}
		\caption{\footnotesize }
		\label{fig:Fig3a}
	\end{subfigure}%
	\begin{subfigure}{.3\textwidth}
		\centering
		\includegraphics[width=0.78\linewidth]{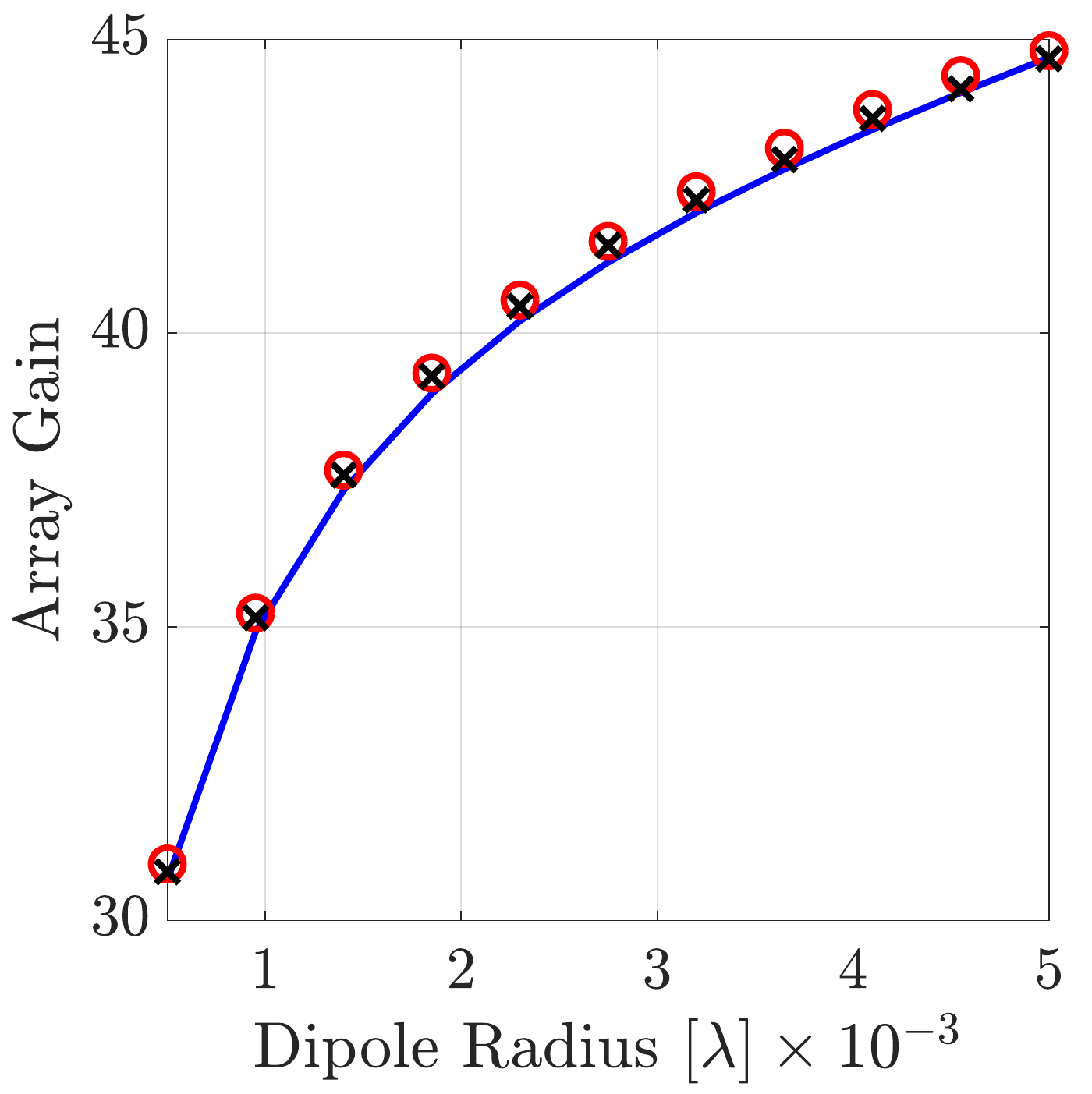}
		\caption{\footnotesize}
		\label{fig:Fig3b}
	\end{subfigure}%
\begin{subfigure}{.3\textwidth}
	\centering
	\includegraphics[width=0.78\linewidth]{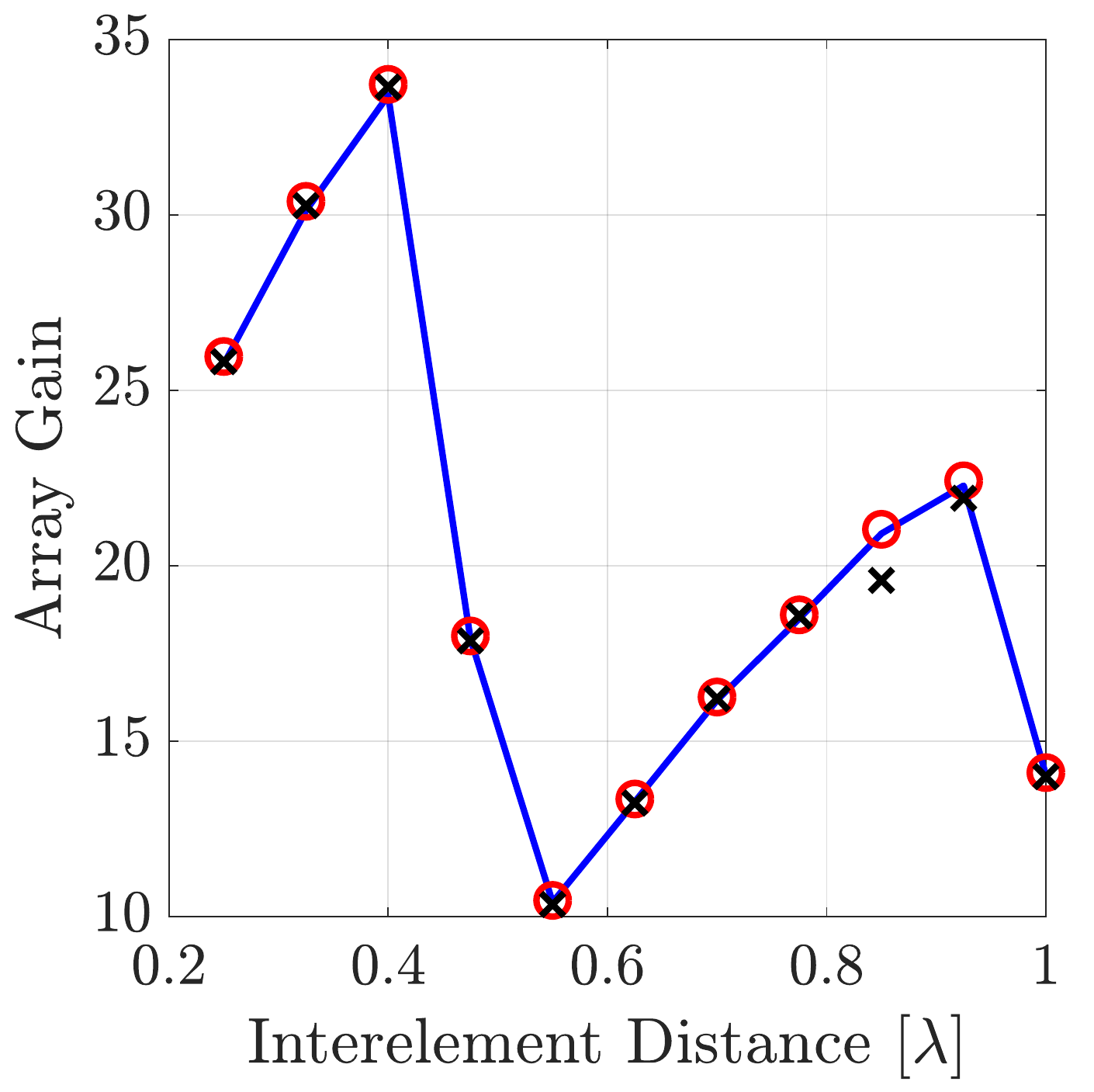}
	\caption{\footnotesize }
	\label{fig:Fig3c}
\end{subfigure}
	\caption{$G_{\max}(\pi/2,0)$ for an endfire \ac{ULA} with $N=10$ elements. The antennas are made of copper with conductivity $\sigma = 5.7 \times 10^7$ S/m, and are placed along the $x$-axis. In the \ac{MoM}-based approach and full-wave simulation, $2M+1=401$ samples have been used. In (a), (b), and (c), $\rho=\lambda/2000$, $\ell = \lambda/2$, and $d =\lambda/3$, while the respective parameter varies accordingly.}
	\label{fig:Fig3}
\end{figure*}

\subsection{Array Model Validation}\label{sec:model_valication}
\begin{figure*}[t]
	\centering
	\begin{subfigure}{.3\textwidth}
		\centering
		\includegraphics[width=0.99\linewidth]{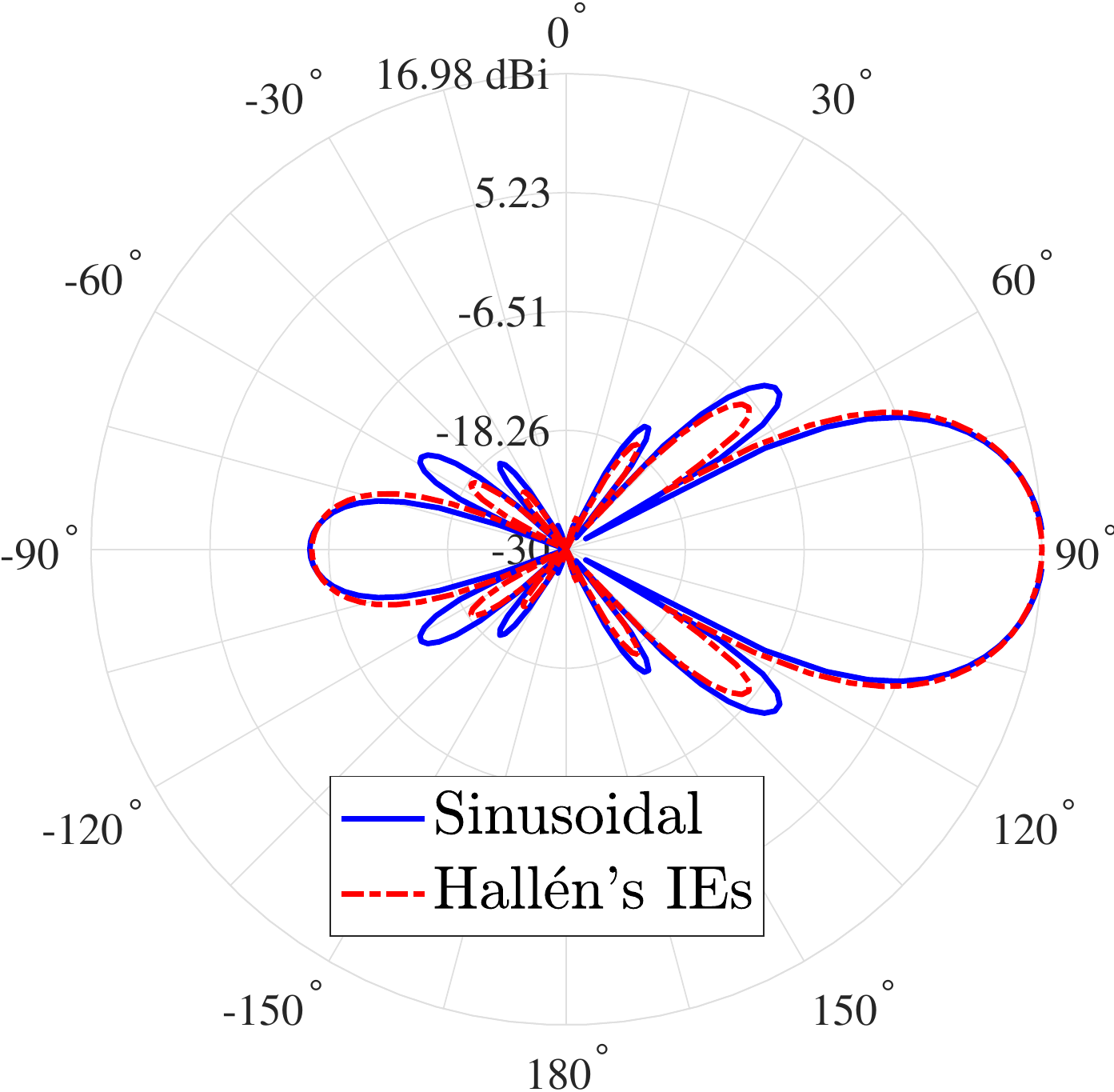}
		\caption{\footnotesize 2D pattern.}
		\label{fig:Fig4a}
	\end{subfigure}\hspace{4cm}
	\begin{subfigure}{0.3\textwidth}
		\centering
		\includegraphics[width=0.99\linewidth]{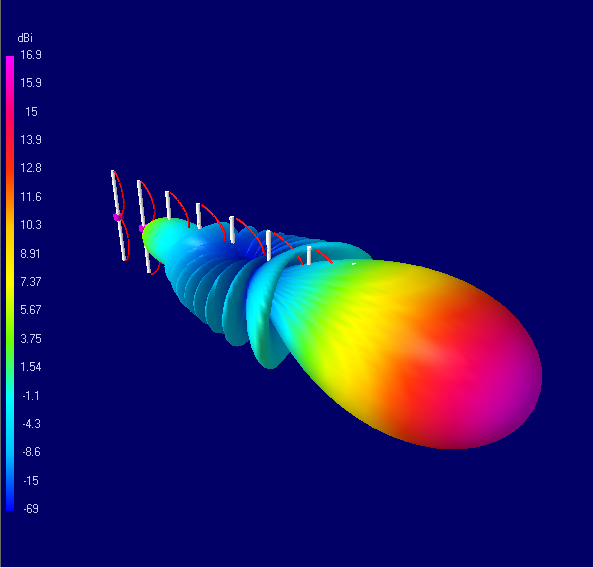}
		\caption{\footnotesize 3D pattern with 4NEC2.}
		\label{fig:Fig4b}
	\end{subfigure}%
	\caption{$G(\theta,\phi)$ for an $10$-element ULA along the $x$-axis at $f=10$ GHz, $d=\lambda/2.5$, and $\mathbf{i}$ given by~\eqref{eq:optimal_bf} for $(\theta,\phi)=(\pi/2,0)$. The dipoles are copper wires of $\ell = 0.9\lambda$ and $\rho = \lambda/200$. In the \ac{MoM} approach and full-wave simulation, $2M+1=401$ samples have been used.}
	\label{fig:Fig4}
\end{figure*}

It is known that the \ac{SCD} is very accurate for dipoles of vanishing radius, i.e., $\rho\to 0$~\cite{balanis_book}. It is therefore important to validate our results for thin dipoles of finite radius, which are closely spaced. For this purpose, we now recall that, under the \textit{thin-wire approximation}, $E_\theta$ takes the form~\cite[Ch. 25]{orfanidis_book} 
\begin{align}\label{eq:e_field2}
E_\theta = j Z_0 k\frac{e^{-jkr}}{4\pi r}\sin\theta\sum_{n=0}^{N-1}e^{jk\hat{\mathbf{r}}\cdot\mathbf{r}_n}S_n(\theta),
\end{align} 
where
\begin{equation}\label{eq:space_factor}
S_n(\theta) \triangleq \int_{-\ell/2}^{\ell/2}I_n(z')e^{jk z'\cos\theta}\text{d}z'
\end{equation}
is the space factor of the $n$th dipole, i.e., line source of length~$\ell$. The current distribution $I_n(z')$ on the $n$th antenna is the result of the driving voltages and their mutual interaction. Thus, the current distributions satisfy a system of coupled Hall\'{e}n's \ac{IEs}, which effectively capture the electromagnetic coupling between adjacent antennas.\footnote{Note that Hall\'{e}n's \ac{IEs} hold for delta-gap input voltages.} These IEs are numerically solved by the \ac{MoM} to obtain $\{I_n(z')\}_{n=0}^{N-1}$ for \textit{given input voltages}. To this end, the basis expansion 
\begin{equation}\label{eq:current_expansion}
I_n(z') = \sum_{m-M}^M I_n(m\Delta)B(z'-m\Delta)
\end{equation}
is employed, where $B(\cdot)$ is a basis function, $2M+1$ is the total number of samples, whereas $\Delta = \ell/(2M)$ is the sample spacing. Considering the pulse basis function
\begin{equation}\label{eq:basis_function}
B(z'- m\Delta) = \left\{
\begin{array}{l l}
1, & \quad |z' - m\Delta | \leq \frac{\Delta}{2}, \\
0,& \quad \text{otherwise}
\end{array}
\right. ,
\end{equation}
the space factor is recast as~\cite[Ch. 25]{orfanidis_book}
\begin{equation}\label{eq:space_factor_sampled}
S_n(\theta)= \sum_{m=-M}^{M} I_n(m\Delta)e^{jkm\Delta\cos\theta}\frac{\sin(k/2\cos\theta \Delta)}{k/2\cos\theta}.
\end{equation}
Based on~\eqref{eq:e_field2} and~\eqref{eq:space_factor_sampled}, the electric field of the $N$ coupled dipoles can be precisely characterized. Next, the radiation intensity becomes
\begin{equation}\label{eq:radiation_intensity_ie}
U =  \frac{Z_0k^2}{32\pi^2}\left|\sum_{n=0}^{N-1}e^{jk\hat{\mathbf{r}}\cdot\mathbf{r}_n}S_n(\theta)\right|^2,
\end{equation}
whilst the radiated power is conveniently computed as
\begin{equation}\label{eq:radiated_power_ie}
P_{\text{rad}} = \frac{1}{2}\text{Re}\left\{\mathbf{v}_{\text{in}}^H\mathbf{i}_{\text{in}}\right\},
\end{equation} 
where $\mathbf{i}_{\text{in}} \triangleq [I_0(0),\dots, I_{N-1}(0)]^T\in\mathbb{C}^{N\times 1}$ and $\mathbf{v}_{\text{in}}\in\mathbb{C}^{N\times 1}$ are the vectors of input currents and voltages, respectively. Lastly, the power loss due to heat dissipation at the $n$th dipole~is 
\begin{equation}
P_{\text{loss},n} \triangleq \frac{1}{2}\bar{R}_{\text{loss}}\!\int_{-\ell/2}^{\ell/2}|I_n(z')|^2 \text{d}z' = \frac{1}{2}\bar{R}_{\text{loss}}\sum_{m-M}^M \! \! |I_n(m\Delta)|^2\Delta,
\end{equation}
which results in the overall power loss
\begin{equation}\label{eq:power_loss_ie}
P_{\text{loss}} =  \sum_{n=0}^{N-1}P_{\text{loss},n}. 
\end{equation}
The following algorithm describes the steps to evaluate the array gain using the electric field IEs for coupled dipoles.
\begin{algorithm}[H]
	\renewcommand\thealgorithm{}
	\caption{Array Gain based on the \ac{MoM}}
	
	\begin{algorithmic}[1]
		\State Assume sinusoidal distribution and calculate the antenna current vector $\mathbf{i}$ using~\eqref{eq:optimal_bf}.
		\State Specify the input voltages as $\mathbf{v}_{\text{in}} = \mathbf{Z}\mathbf{i}$, where $\mathbf{Z}$ is computed by the induced EMF method for lossless antennas and~\ac{SCD}.
		\State For given $\mathbf{v}_{\text{in}}$, obtain $\{I_n(z')\}_{n=0}^{N-1}$ from Hall\'{e}n's \ac{IEs}.
		\State Compute $G(\theta,\phi)  =  4\pi U /P_{\text{in}}$ using~\eqref{eq:radiation_intensity_ie},~\eqref{eq:radiated_power_ie} and~\eqref{eq:power_loss_ie}.
	\end{algorithmic}\label{algo1}
\end{algorithm} 
\begin{remark}
The \ac{SCD} renders the impedance matrix $\mathbf{Z}$ independent of the input voltages~\cite{orfanidis_book}. As a result, the entries of $\mathbf{Z}$ hinge solely on the array geometry and antenna characteristics. This facilitates the computation of $\mathbf{Z}$, which is used to theoretically determine the optimal current excitation via~\eqref{eq:optimal_bf}. 
\end{remark}
We now examine the accuracy of the \ac{SCD} assumption when calculating the array gain $G_{\max}(\pi/2,0)$ for different dipole lengths. From Fig.~\ref{fig:Fig3}(\subref{fig:Fig3a}), we first confirm the excellent match between the theoretical model, the \ac{MoM}-based approach, and the full-wave simulation. The small discrepancy at $\ell = 0.9\lambda$ is expected, because the \ac{SCD} assumption breaks down as the dipole length approaches $\lambda$~\cite{balanis_book}. From Fig.~\ref{fig:Fig3}(\subref{fig:Fig3b}), we also see a good agreement for various dipole radii. More importantly, this holds for dipoles as thick as $\rho=\lambda/200$. Regarding the interelement spacing, 
the optimal one is $\lambda/2.5$ according to Fig.~\ref{fig:Fig3}(\subref{fig:Fig3c}), which implies that the dipoles should not be placed very close to each other; similar finding were reported in~\cite{he_sg_arrays}, though for isotropic radiators. Lastly, Fig.~\ref{fig:Fig4} depicts the 2D and 3D gain patterns under optimal interelement separation, which were calculated using the \ac{SCD} assumption, Hall\'{e}n's \ac{IEs}, and full-wave simulation. As expected, the maximum array gain is achieved along the endfire direction $(\pi/2,0)$, and is $16.98$ dBi (i.e., 49.88 in linear scale). In conclusion, the proposed array model can provide meaningful results, yet with much smaller computational complexity than purely numerical methods. 

\section{Conclusions}
We studied, for the first time, the impact of dipole antenna dimensions on superdirectivity. For this purpose, we developed an array model that captures the main characteristics of linear dipoles. Capitalizing on the \ac{SCD} of very thin wires, the overall ohmic losses were explicitly computed, which greatly affect the array gain. Next, the optimal beamforming problem under a fixed power constraint was addressed. As shown, a super-gain can be attained without sacrificing the energy efficiency of the system when not too short and thin elements are employed. We also confirmed our findings via a \ac{MoM}-based approach as well as full-wave simulations. In particular, it was demonstrated that the proposed theoretical model predicts the array gain of coupled thin dipoles with high precision.  

\section*{Acknowledgements}
This project has received funding from the European Research Council (ERC) under the European Union’s Horizon 2020 research and innovation programme (grant agreement No. 101001331).

\end{document}